\documentclass[preprint, showpacs, preprintnumbers ,amsmath,amssymb]{revtex4}

\usepackage{graphicx}
\usepackage{dcolumn}
\usepackage{bm}

\begin{document}




\title{ Electron-hydrogen excitation to the $n=3$ and $n=4$ levels in the Glauber approximation}


\author{Victor Franco}

\affiliation{Physics Department, Brooklyn College of the City
University of New York, Brooklyn, New York, 11210}

\date{\today}

\begin{abstract}
We have calculated the differential and integrated cross sections
for excitation of atomic hydrogen to its $n=3$ and $n=4$ levels by
electron impact using the Glauber approximation.  Results are
compared with measurements at $20, 30, 40$, and $60$ eV and also
shown for $120$ and $480$ eV.  At momentum transfers not too large
at all energies considered, the calculated $n=3$ differential
cross sections are qualitatively similar to but  a factor of
somewhat less than  $3$ larger than the calculated $n=4$ cross
sections. The calculated integrated cross sections attain broad
maxima near $41$ eV.

\end{abstract}


\maketitle


    Only relatively recently have considerably accurate theoretical
     methods for describing electron-hydrogen atom scattering been
     presented \cite{1,2,3,4}.  These methods require very-large-scale
      calculations or involved computations such as six-dimensional
       numerical quadrature.  It would be useful to be able to calculate
        electron-hydrogen cross sections with ease without sacrificing
        too much of the accuracy of the established reliable methods.
        The first-Born approximation is perhaps the most easily utilized
        scattering approximation, but its validity is questionable except
         at rather high incident energies.  And furthermore,
         the convergence of the Born approximation to the correct
         nonrelativistic result may occur only at such high energies
          that relativistic effects become significant \cite{5}. \\

    An approximation which has been used extensively with much success
    in nuclear  physics (and with moderate success in atomic physics)
    is the Glauber approximation \cite{6}.  This approximation
    is expected to be valid at high incident energies and small
    momentum transfers.  It was introduced to atomic physics in $1968$ \cite{7}, and
     is  as easy to use today as the Born approximation was then.
      With a laptop personal computer, complete numerical calculations,
      in most cases, take of the order of a second.\\

    The Glauber approximation amplitude, $f(1s \rightarrow nlm;\textbf{q})$, for excitation
    of the ($nlm$) level of atomic hydrogen by electrons with incident momentum
    $\hbar {\bf k}$ and final momentum $\hbar {\bf k}_{f}$ is given by Thomas and Franco \cite{8}.
      It may be rewritten as the explicit closed-form expression
      \begin{eqnarray}
f(1s \rightarrow nlm;\textbf{q}) &=&\frac{(-1)^{l+1} 2^{2l+4}
\sqrt{(n-l)_{2l+1}} }{i^{|m|} n^{l+2} (2l+1)!} \sqrt{\pi} \;
{Y_{l}^{m}}^{*}(\frac{\pi}{2}, \varphi_{q})  \nonumber \\
& \times &
 \frac{\Gamma(1+i \eta)\Gamma[\frac{1}{2}(l+|m|)-i \eta]\Gamma(1+|m|-i \eta)}{\Gamma(1-i
 \eta)|m|!} \nonumber \\
 & \times &
 \frac{1}{(a_{0}q)^{l+4}} \sum_{j=0}^{n-l-1} \frac{(-n+l+1)_{j}}{j!
 (2l+2)_{j}}\left(\frac{2}{n a_{0}}\right)^{j} \left(\frac{\partial}{\partial
 \lambda}\right)^{j+1} \left(\frac{\partial}{\partial z}\right)^{1+(l-|m|)/2} \nonumber \\
& \times &  \{z^{-i \eta}{}_{2}F_{1}[ (l+|m|)/2-i\eta;
1+|m|-i\eta; 1+|m|; -z]
 \}
\end{eqnarray}
where $z =\lambda^{2}/q^{2}$ with $\lambda$ to be evaluated at
$\lambda=(1+1/n)a_{0}$. Here $\hbar {\bf q}=\hbar {\bf k}-\hbar
{\bf k}_{f}$ is the momentum transfer, $a_{0}$ is the Bohr radius,
$(a)_{j}$ is Pochhammer's symbol \cite{9}, and $\eta=e^{2}/\hbar
\nu$ with $\nu$ being the incident electron speed. The
corresponding differential cross section is given by
\begin{equation}
d\sigma/d\Omega=(k_{f}/k)\left|f(1s \rightarrow
nlm;\textbf{q})\right|^{2}.
\end{equation}

Equation $(1)$ shows that the cross section is given in terms of a
linear combination of products of (complex) powers of $q$ and
derivatives of hypergeometric functions.  If one uses the
differentiation and recursion relations satisfied by the
hypergeometric function \cite{9}, the full result may be expressed
in terms of only two contiguous hypergeometric functions and
simple functions of $q$.  No advantage is gained, however, by
explicitly exhibiting the results in that form since the algebra
involved is very tedious and the forms of the simple functions of
$q$ are quite complicated, lengthy, and not at all transparent.
It is more useful to express the results directly in terms of
 derivatives with respect to $z$ of the
$z^{-i\eta}{}_{2}F_{1}$ function in Eq. $(1)$.
      Closed form expressions in terms of such derivatives are then
      obtained.\\

Equations (1) and (2) may be used to calculate cross sections for
excitation of ground state atomic hydrogen to any excited state.
We present the explicit results for excitation to the $n=3$ and
$n=4$ levels.\\

The cross section for excitation to the $n=3$ level is given by
\begin{eqnarray}
\frac{d\sigma}{d \Omega}(1s \rightarrow n=3; {\bf q})
&=&\frac{k_{f}}{k} \frac{3^{7}}{2^{16}} a_{0}^{2}  \left(
\frac{\pi \eta }{ sinh(\pi \eta)} \right)^{2} z^{6}  \{
\frac{z^{2}}{18} \left| \frac{\partial^{3}}{\partial z^{3}} [z^{-i
\eta}{}_{2}F_{1}(1-i \eta, 1-i \eta;1; -z)]\right|^{2}
\nonumber \\
&+&  \frac{z^{2}}{24}  (4+\eta^{2}) (1+\eta^{2})^{2}
\left|\frac{\partial^{2}}{\partial z^{2}}[z^{-i \eta}
{}_{2}F_{1}(2-i \eta, 3-i \eta;3;-z)]\right|^{2} \nonumber \\
&+ &  \frac{3 z}{2}(1+\eta^{2})\left|\left(3
\frac{\partial^{2}}{\partial z^{2}} + \frac{2}{3} z
\frac{\partial^{3}}{\partial z^{3}} \right) [z^{-i \eta }
{}_{2}F_{1}(1-i \eta, 2-i \eta; 2; -z)]\right |^{2} \nonumber \\
&+ & \frac{1}{\eta^{2}}\left | \left(6
\frac{\partial^{2}}{\partial z^{2}} + 5 z \frac{\partial^{3}
}{\partial z^{3} }+ \frac{2}{3}z^{2} \frac{\partial^{4}}{\partial
z^{4}} \right) [z^{-i \eta } {}_{2}F_{1}(-i \eta, 1-i \eta; 1;
-z)]\right|^{2} \}
\end{eqnarray}
where the right hand side is to be evaluated at
$z=16/[9(a_{0}q)^{2}]$.\\

    The cross section for excitation to the $n=4$ level is given by

\begin{eqnarray}
\frac{d\sigma}{d \Omega}(1s &\rightarrow &n = 4; {\bf q})  =
\frac{k_{f}}{k} \frac{2^{27}}{5^{16}} a_{0}^{2} \left(\frac{\pi
\eta}{sinh(\pi \eta)} \right)^{2} z^{6}
\{\frac{z^{3}}{1296}(9+\eta^{2})(4+\eta^{2})^{2}(1+\eta^{2})^{2}
  \nonumber \\
&\times  &   \left| \frac{\partial^{2}}{\partial z^{2}} [z^{-i
\eta }{}_{2}F_{1}(3-i \eta, 4-i \eta;4; -z)] \right|^{2} \nonumber \\
& + & \frac{z^{3}}{60}(1+\eta^{2})^{2}
\left|\frac{\partial^{3}}{\partial z^{3}}
[z^{-i \eta} {}_{2}F_{1}(2-i \eta, 2-i \eta;2;-z)]\right|^{2} \nonumber \\
&+ &  \frac{2 z^{2}}{9}\left| \left(4 \frac{\partial^{3}}{\partial
z^{3}} + \frac{z}{2}\frac{\partial^{4}}{\partial z^{4}}\right)
[z^{-i \eta }
{}_{2}F_{1}(1-i\eta, 1-i \eta; 1; -z)]\right|^{2}  \nonumber \\
&+ & \frac{z^{2}}{6}(4+\eta^{2})(1+\eta^{2})^{2} \left|(4
\frac{\partial^{2}}{\partial z^{2}} + \frac{z}{2}
\frac{\partial^{3}}{\partial z^{3} })[ z^{-i \eta }
{}_{2}F_{1}(2-i\eta, 3-i \eta; 3; -z) ]\right|^{2}\nonumber \\
&+& \frac{5z}{2}(1+\eta^{2}) \left| \left(15\frac{
\partial^{2}}{\partial z^{2}}+\frac{28 z}{5}\frac{
\partial^{3}}{\partial z^{3}}+ \frac{2 z^{2}}{5}\frac{  \partial^{4}}{ \partial z^{4}} \right) [z^{-i \eta }
{}_{2}F_{1}(1-i \eta, 2-i \eta; 2; -z)]\right|^{2} \nonumber \\
&+& \frac{2}{\eta^{2}}\left|\left( 25\frac{\partial^{2}}{\partial
z^{2}}+\frac{53z}{2} \frac{\partial^{3}}{\partial z^{3}}+6 z^{2}
\frac{\partial^{4}}{\partial z^{4}}+ \frac{z^{3}}{3}
\frac{\partial^{5}}{\partial z^{5}}\right) [z^{-i \eta }
{}_{2}F_{1}(-i \eta, 1-i \eta; 1; -z)]\right|^{2}\}
\end{eqnarray}
where the right hand side is to be evaluated at
$z=25/[16(a_{0}q)^{2}]$.\\

    These relatively simple explicit analytic expressions
     may be used to calculate the differential cross sections
     for excitation of ground state hydrogen atoms to the $n=3$ and $n=4$ states.
       As can be seen, each may be found by writing  a single (albeit somewhat lengthy)
        statement.  Using \emph{\textbf{Mathematica}}, for example,
        the entire calculation may be done with one few-line
        input statement and the time required on a PC for the entire differential cross section is of the order of a second.\\

             We have calculated the differential cross sections for the $1s \rightarrow n=3$ and
             $1s \rightarrow n=4$ excitations using Eqs.(2-4) at
             energies of $20$ eV, $30$ eV, $40$ eV, and $60$ eV, where measurements have
              been recently made \cite{10}.  We have also calculated these cross sections at
              $120$ eV and $480$ eV, where the theory has greater
              validity.  Since the theory is valid at high energies and
               small momentum transfers, one would not expect accurate results for
                large momentum transfers (large scattering angles).  However,
                 at high energies the cross sections decrease rapidly from their
                  values near the forward direction and the bulk of the scattering occurs
                   at small momentum transfers.\\

    At energies as low as $20-60$ eV use of the Glauber
    approximation would  not be justified if one required highly
    accurate results.  But even at these low energies, one might expect to
      obtain qualitative results at the lower momentum transfers.  We
      present the results of our calculations for momentum transfers of $q \leq 1.24 k_{f}$, and
      our comparison of the theory with the measurements is exhibited  in that
      domain.  Larger momentum transfers are beyond the range of validity of the
       theory (which does not compare well with the large-$q$ data \cite{10}, as
       expected).\\

\begin{figure}
 \begin{center}
 \includegraphics[bb=0 0 511 521, angle=0.3, scale=0.7]{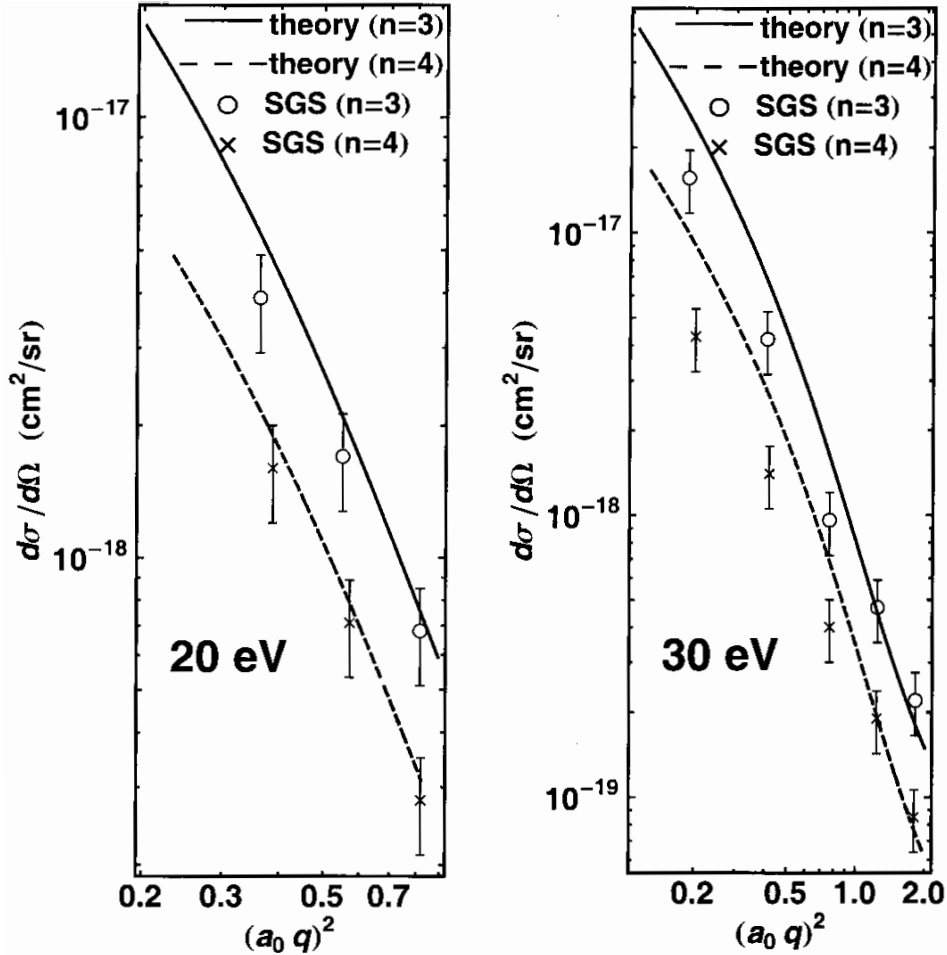}
 \caption{Differential cross sections, as functions of squared momentum transfer, for excitation of
 the $n=3$ and $n=4$ levels of atomic hydrogen by electron impact at $20$ eV and $30$ eV. ``theory'' represents the
 Glauber approximation, Eq.(3) and (4). Also shown are the measurements of Sweeney, Grafe, and Shyn (denoted SGS).\label{}}
 \end{center}
 \end{figure}

\begin{figure}
 \begin{center}
 \includegraphics[bb=0 0 511 521, angle=0.3, scale=0.7]{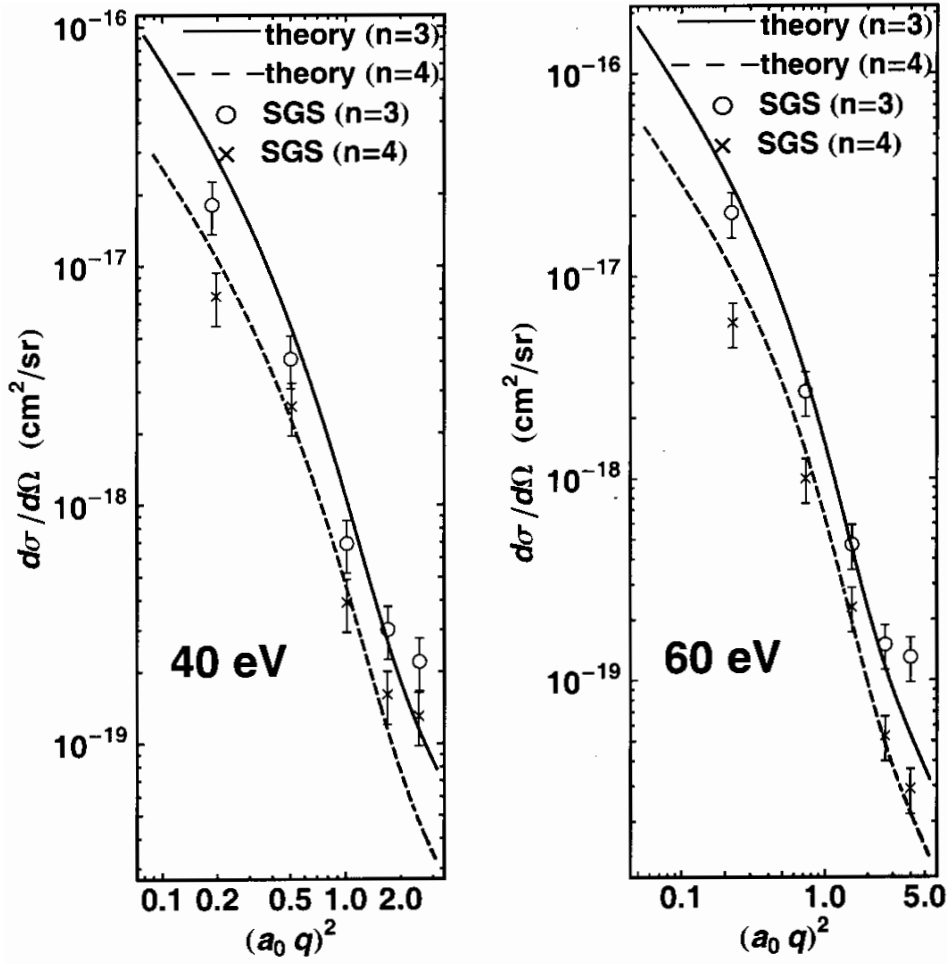}
 \caption{Differential cross sections,  as functions of squared momentum transfer, for excitation of
 the $n=3$ and $n=4$ levels of atomic hydrogen by electron impact at $40$ eV and $60$ eV. See Fig.1 for more details.\label{}}
 \end{center}
 \end{figure}
In Fig.1 and Fig.2 we show the calculated differential cross
sections for $1s \rightarrow  n=3$ and $1s \rightarrow n=4$
excitation as a function of $(a_{0} q)^{2}$ and compare our
results with the measurements (SGS) \cite{10} which, for the three
highest energies, decrease by factors of as much as $50$ to $200$
from their values at the smallest measured momentum transfers. Our
results are in reasonable agreement with the data, even at these
relatively low energies. Both the theory and the measurements
indicate that the $n=3$ and $n=4$ differential cross sections are
qualitatively very similar, with the former being a factor of
 somewhat less than $3$ larger than the latter.  At larger
momentum transfers (not shown), where the theory is not valid, the
calculated cross sections continue to decrease rapidly.  Where
measurements have been made at larger momentum transfers $(1s
\rightarrow n=3$ at all four energies and $1s \rightarrow n=4$ at
$20$ eV and $30$ eV), that is not the case \cite{10}.\\

\begin{figure}
 \begin{center}
 \includegraphics[bb=0 0 510 516, angle=0.3, scale=0.7]{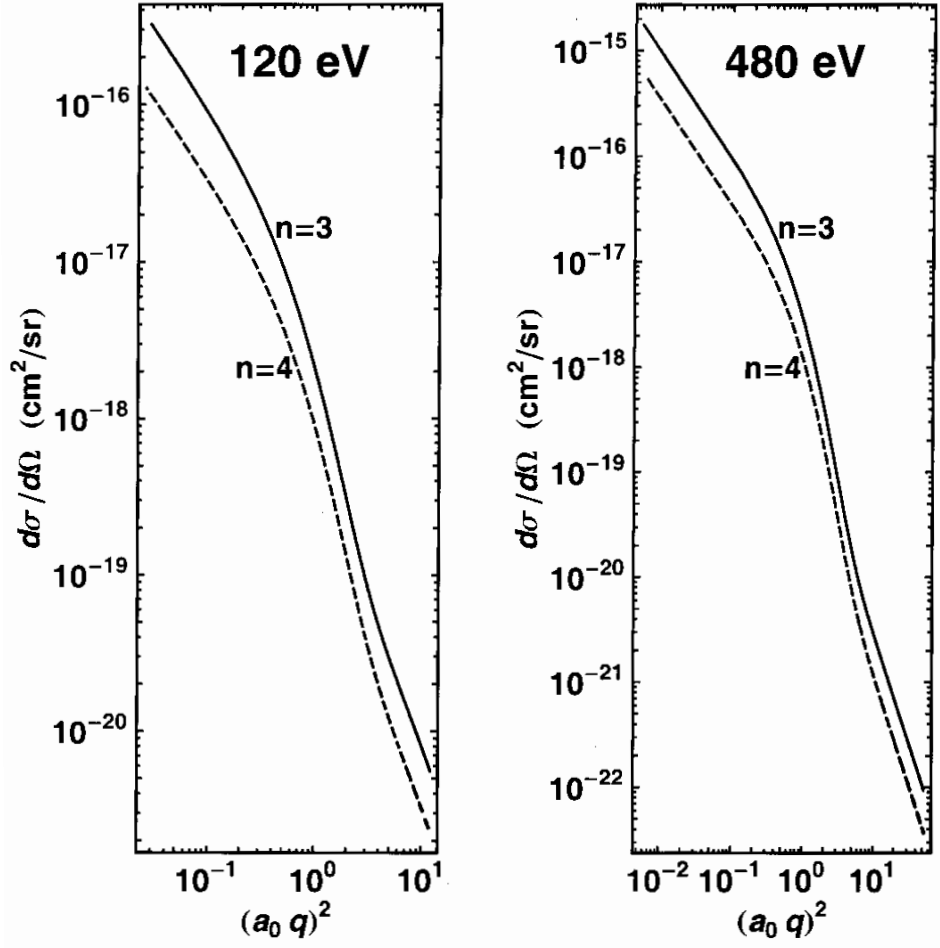}
 \caption{Differential cross sections, as functions of squared momentum transfer, for excitation of
 the $n=3$ and $n=4$ levels of atomic hydrogen by electron impact at $120$ eV and $480$ eV in the Glauber approximation .\label{}}
 \end{center}
 \end{figure}
In Fig.3 we show the calculated results at energies of $120$ eV
and $480$ eV, where the approximation has greater validity.  These
results exhibit the same similarities of the $n=3$ and $n=4$ cross
sections as those exhibited at the lower energies, with the $n=3$
cross sections being a factor of somewhat less than $3$ larger
than the $n=4$ cross sections throughout the entire domain of
momentum transfers shown.\\

The integrated cross sections, obtained by integrating
$d\sigma/d\Omega$ over $d\Omega$, are calculated to be (in units
of $10^{-18}$ $cm^{2}$): $10.0$, $14.4$, $15.4$, $14.5$, $9.90$,
$6.79$, and $4.08$ for $1s \rightarrow n=3$, and $3.55$, $5.28$,
$5.64$, $5.31$, $3.86$, $2.45$, and $1.47$ for $1s \rightarrow
n=4$, at $20, 30, 40, 60, 120, 240$, and $480$ eV, in that order.
These results may be compared with measured values \cite{10} of
$11.4\pm3.1$  and $10.9 \pm 2.9$ for $1s \rightarrow n=3$ at $20$
eV and $30$ eV, respectively, and $5.28\pm 1.43$ for $1s
\rightarrow n=4$ at $20$ eV. The ratios of the calculated $n=3$ to
$n=4$ cross sections are  approximately constant  ($2.7$).  The
lone measurement of this ratio is $2.2\pm 0.8$ at $20$ eV and the
calculated value is $2.8$. The calculated $n=3$ and $n=4$ cross
sections attain broad maxima of $15.4$ and $5.64$, respectively,
both near $41$ eV.\\

In conclusion, we have presented explicit and easily calculable
closed-form  expressions for the differential cross sections for
excitation of atomic hydrogen to its $n=3$ and $n=4$ levels in
terms of derivatives of hypergeometric functions.  The results are
in qualitative agreement with measurements at the relatively low
energies of $20-60$ eV at momentum transfers that are not too
large. Calculations at these energies as well as those presented
for higher incident energies, where the theory has greater
validity, indicate qualitative similarity between the $n=3$ and
$n=4$ cross sections, with the n=3 cross
sections being  larger by a factor of somewhat less than $3$. This is also
exhibited in the calculated integrated cross sections, which attain broad maxima near $41$ eV.\\


\begin{references}
\bibitem{1}I. Bray and A. T. Stelbovics, Phys. Rev. A \textbf{46}, 6995 (1992);
 I. Bray, Phys. Rev. A\textbf{ 49},  1066 (1994);
I. Bray, J. Phys. B \textbf{33}, 581 (2000); I. Bray, Aust. J.
Phys. \textbf{53}, 355 (2000).

\bibitem{2}
M. Baertschy, T. N. Resigno, W. A. Isaacs, and C. W. McCurdy,
Phys. Rev. A \textbf{60},  R13 (1999); T. N. Resigno, M.
Baertschy, W. A. Isaacs, and C. W. McCurdy, Science \textbf{286},
2475 (1999).

\bibitem{3} D. S. F. Crothers and J. F. McCann, J. Phys. B\textbf{ 16}, 3229 (1983).


\bibitem{4} S. Jones and D. H. Madison, Phys. Rev. Lett. \textbf{81}, 2886 (1998).

\bibitem{5} S. Jones and D. H. Madison, Phys. Rev. A \textbf{66}, 062711 (2002).

\bibitem{6} R. J. Glauber, in \emph{Lectures in Theoretical Physics}, edited by W. E. Brittin
 et al. (Interscience, New York, 1959), Vol. I, p.315.



\bibitem{7} V. Franco, Phys. Rev. Lett. \textbf{20}, 709 (1968).

\bibitem{8} B. K. Thomas and V. Franco, Phys. Rev. A \textbf{13}, 2004 (1976).




\bibitem{9}W. Magnus, F. Oberhettinger, and R. P. Soni,
\emph{Formulas and Theorems for the Special Functions of
Mathematical Physics} (Springer-Verlag, New York, 1966).


\bibitem{10}C. J. Sweeney, A. Grafe, and T. W. Shyn, Phys. Rev A \textbf{64}, 032704 (2001);
 C. J. Sweeney, A. Grafe, and T. W. Shyn, Phys. Rev. A \textbf{69}, 052709 (2004).





\end{references}
\end{document}